\documentclass[a4paper,12pt]{article}
\usepackage{amssymb,amsmath}
\usepackage{bm}
\usepackage{cite}

\def\beq{\begin{equation}}
\def\eeq{\end{equation}}
\def\bea{\begin{eqnarray}}
\def\eea{\end{eqnarray}}
\def\nn{\nonumber}

\def\sch#1{\hat{\mathfrak{s}}(#1)}
\def\del#1{\partial_{#1}}
\def\D{{\Delta}}
\def\hX{{\hat X}}
\def\tX{{\tilde X}}
\def\tL{{\tilde L}}
\def\a{{\alpha}}
\def\bbz{Z\!\!\!Z}

\begin{document}
\pagestyle{empty}

\begin{center}

 \textsf{\LARGE Intertwining Operator Realization of\\[2mm]
Non-Relativistic Holography}

\vspace{7mm}

{\large N.~Aizawa$^{a,}$\footnote{aizawa@mi.s.osakafu-u.ac.jp}, ~~~
V.K.~Dobrev$^{b,}$\footnote{dobrev@inrne.bas.bg}
 }

\vspace{3mm}

 \emph{$^a$ Department of Mathematics and Information Sciences,\\ Graduate School of Science,
Osaka Prefecture University, Nakamozu Campus,\\ Sakai, Osaka 599-8531 %%% 590-0035,
Japan} \vspace{3mm}

 \emph{$^b$Institute of Nuclear Research and Nuclear Energy,\\
 Bulgarian Academy of Sciences \\
72 Tsarigradsko Chaussee, 1784 Sofia, Bulgaria}

\end{center}

\vspace{.8 cm}

\begin{abstract}
We give a group-theoretic interpretation of non-relativistic
holography as equivalence between representations of the
Schr\"odinger algebra describing  bulk fields  and  boundary fields.
Our main result is the explicit construction of the boundary-to-bulk
operators
%%% NA Aug 25  from here
%for which we first construct the two-point Green function
%in the bulk.
in the framework of representation theory %%%%
(without specifying any action).
%%% to here
Further we show that these operators and the bulk-to-boundary
operators are intertwining operators. In analogy to the relativistic
case, we show that each bulk field has two boundary fields with
conjugated conformal weights. These fields are related by another
intertwining operator given by a two-point function on the boundary.
Analogously to the relativistic result of Klebanov-Witten we give
the conditions when both boundary fields are physical. Finally, we
recover in our formalism %%%%
earlier non-relativistic results
%%% NA Aug 25
%in a less general setting by Son and others.
for scalar fields by Son and others.%%%%
\end{abstract}

\vfill\newpage \pagestyle{plain} \setcounter{page}{1}

\section{Introduction}

The role of nonrelativistic symmetries in string theory was always
important. In fact, being the theory of everything string theory
encompasses together relativistic quantum field theory, classical
gravity, and certainly, non-relativistic quantum mechanics, in such
a way that it is not even necessary to separate these components,
cf., e.g., \cite{MaMaTa,AoNiSu,Schvellinger,AADV}.

  Thus, it is not a surprise that the
Schr\"odinger group - the group that is the maximal group of
symmetry of the Schr\"odinger equation - is playing more and more a
prominent role, cf., e.g.,
\cite{NisSon,Son,BaMcG,Goldberger,MinPle,DuHaHo,Taylor,AMSV,DonGau,BagGop,ColYava,BoKuPi,OogPar}.

Originally, the Schr\"odinger group, actually the Schr\"odinger
algebra, was introduced by Niederer \cite{Nie} and Hagen \cite{Ha},
as a nonrelativistic limit of the vector-field realization of
the conformal algebra. In the process, the space components of
special conformal transformations decouple from the system. Thus,
e.g., in the case of four-dimensional Minkowski space-time from the
15 generators of the conformal algebra we obtain the 12 generators
of the Schr\"odinger algebra.

Recently, Son \cite{Son} proposed another method of identifying the
Schr\"odinger algebra in d+1 space-time. Namely, Son started from
AdS space in d+3 dimensional space-time with metric that is
invariant under the corresponding conformal algebra  so(d+1,2)  and
then deformed the AdS metric to reduce the symmetry to the
Schr\"odinger algebra.

In view of the relation of the conformal and Schr\"odinger algebra
there arises the natural question.  Is there a nonrelativistic
analogue of the AdS/CFT correspondence,
in which the conformal
symmetry is replaced by Schr\"odinger symmetry.  Indeed, this is to
be expected since the Schr\"odinger equation should play a role both
in the bulk and on the boundary.  The posed question was studied in
some of the literature above, and also in
\cite{BroTer,BarFue,HeRaRo,AdBaMcG,Yamada,SakYos,RSST,FueMor,VolWen}.
%%% NA Aug 25 from here
%However, the typical approach is to treat the boundary as
%one-dimensional, cf. e.g. \cite{BaMcG}. Furthermore it seems that
%the remaining variable is exactly the variable that distinguishes
%the bulk from the boundary in the standard AdS/CFT approach
%\cite{Maldacena,GuKlPo,Witten}.
%
%We think it would be more consistent to  implement the standard
%holographic picture in which the boundary has  the naturally
%expected dimension of just one dimension less than the bulk.
%
%This is what we do in the present paper explicitly for the
%$(3+1)$-dimensional bulk.

  In the present paper, we examine the nonrelativistic analogue of
the AdS/CFT correspondence in the framework of representation
theory.%%%%
Before explaining what we do let us remind
that the AdS/CFT correspondence %%%% Is this "which" necessary ? question by NA
has 2 ingredients
\cite{Maldacena,GuKlPo,Witten}: %%% NA Aug 25
1. the holography principle, which is very
old, and  means the reconstruction of some objects in the bulk (that
may be classical or quantum) from some objects on the boundary; 2.
the reconstruction of quantum objects, like 2-point functions on the
boundary, from appropriate actions on the bulk.
%%% NA Aug 25 from here
Our main focus is put on the first ingredient and we consider the simplest
case of the (3+1)-dimensional bulk.
It is shown that the holography principle is established using
representation theory only, that is, we do not specify any action.
We outline the contents of the paper below.
%%% to here

For the implementation of the first ingredient in the Schr\"odinger
algebra context we use a method that is used in the mathematical
literature for the construction of discrete series representations
of real semisimple Lie groups \cite{Hotta,Schmid}, and which method
was applied in the physics literature first in \cite{DMPPT} in
exactly an AdS/CFT setting, though that term was not used then.

The method  utilizes the fact that in the bulk the Casimir operators
are not fixed numerically. Thus, when a vector-field realization of
the algebra in consideration is substituted in the Casimir it turns
into a  differential  operator.  In contrast, the boundary Casimir
operators are fixed by the quantum numbers of the fields under
consideration. Then the bulk/boundary correspondence forces an
eigenvalue equation involving the Casimir differential operator.
That eigenvalue equation is used to find the two-point Green
function in the bulk which is then used to construct the
boundary-to-bulk integral operator. This operator maps a boundary
field   to a bulk field similarly to what was done in the conformal
context by Witten (cf., e.g., formula (2.20) of \cite{Witten}).
This is our first main result. %%% NA Aug 25

%%% NA Aug 25
%What is also important in our approach is
Our second main result is that we show that this
operator is an intertwining operator, namely, it intertwines the two
representations of the Schr\"odinger algebra acting in the bulk and
on the boundary. This also helps us to establish that each bulk
field has actually two bulk-to-boundary limits. The two boundary
fields have conjugated conformal weights ~$\Delta$, $3-\Delta$, and
they are related by a  boundary two-point function.

We consider also the second ingredient of the AdS/CFT correspondence
in the Schr\"o\-dinger context
%%% NA Aug 25 from here
%and show how to obtain in our
%formalism the two-point function on the boundary recovering results
%of \cite{Son,BaMcG}. Thus, we note also that our approach is more
%general, since we can reproduce the relevant earlier results.
%Furthermore, we can easily extend our considerations for the
%higher-dimensional cases \cite{AizDob}.
and show how our formalism involving the Casimir differential
operator relates to the case of scalar field theory discussed in
\cite{Son,BaMcG}.
We can easily extend our considerations for the
higher-dimensional cases \cite{AizDob}. Higher dimensional
Schr\"odinger group has the rotation group as a subgroup.
Thus our formalism can be naturally extended to the cases with arbitrary spin.
%%% to here

The paper is organized as follows.  In Section 2 we give the
preliminaries on the Schr\"odinger algebra including the Casimir and
the well-known  vector-field realization. In Section 3 we make the
choice of bulk using the four-dimensional space of Son and write
down the vector-field realization in the bulk. In Section 4 we
construct the integral boundary-to-bulk operator. In Section 5 we
establish the intertwining properties of the boundary-to-bulk and
bulk-to-boundary operators. We display also the intertwining
relation between the two bulk-to-boundary limits of a bulk field.
Finally, in Section 6 we relate our approach to earlier work on
non-relativistic holography showing how we can recover those results
for $d=1$.

\section{Preliminaries}

The Schr\"odinger algebra $ {\mathfrak s}(d) $ in
($d$+1)-dimensional spacetime is generated by time translation
$P_t$, space translation $ P_k$, Galilei boosts $ G_k$, rotations $
J_{k\ell} = -J_{\ell k} $ (which generate the subalgebra $ so(d)$),
dilatation $D$ and conformal transformation $K$ ($k, \ell = 1,
\cdots d$). The non-trivial commutation relations are \cite{BarRac}
\begin{equation}
\begin{tabular}{llll}
 $[P_t, D] = 2P_t, \ $& $ [P_t, G_k] = P_k,  $&$ [P_t, K] = D,$  & $[P_k, D] = P_k, \ $ \\
 $ [P_k, K] = G_k, $ & $ [D, G_k] = G_k,$ & $[D, K] = 2K, $ & \\
 \multicolumn{2}{l}{$[P_i, J_{k\ell}] = \delta_{i\ell} P_k - \delta_{ik} P_{\ell},$} &
 \multicolumn{2}{l}{$[G_i, J_{k\ell}] = \delta_{i\ell} G_k - \delta_{ik} G_{\ell},$} \\
 \multicolumn{3}{l}{$[J_{ij}, J_{k\ell}] = \delta_{ik}J_{j\ell}
 + \delta_{j\ell}J_{ik}
 -\delta_{i\ell}J_{jk} - \delta_{jk}J_{i\ell},$} &
\end{tabular}
\label{commsn}
\end{equation}
Actually, we shall work with the central extension $\hat {\mathfrak
s}(d) $ of the Schr\"odinger algebra obtained by adding the central
element $M$ to ${\mathfrak s}(d)$ which enters the additional
commutation relations: $
 [P_k, G_{\ell}] = \delta_{k\ell} M.
$ In many physical applications the central element $M$ corresponds
to mass.

For the purposes of this paper we now restrict to the 1+1
dimensional case. In this case the centrally extended Schr\"odinger
algebra has six generators:

\medskip
\begin{tabular}{cclcccl}
  $H$ & : & time translation & \qquad & $P$ & : & space translation \\
  $G$ & : & Galilei boost & & $D$ & : & dilatation \\
  $K$ & : & conformal transformation & & $M$ & : & center
\end{tabular}\\
with the following non-vanishing commutation relations: \beq
  \begin{array}{lclcl}
    [H, D] = 2H, & & [D, K] = 2K, & & [H, K] = D, \\[3pt]
    [P, G] = M,  & & [P, K] = G,  & & [H, G] = P, \\[3pt]
    [P, D] = P,  & & [D, G] = G.  & &

  \end{array}
  \label{S1def}
\eeq

For our approach we need the Casimir operator. It turns out that the
lowest order nontrivial Casimir operator is the 4-th order one
 \cite{Perroud}:
\beq
  \tilde{C}_4 = ( 2 M D - \{ P, G \} )^2 - 2 \{ 2MK - G^2, 2MH-P^2 \}.
  \label{Casimir}
\eeq In fact, there are many cancellations, and the central
generator $M$ is a  common linear multiple.\footnote{This is seen
immediately by setting $M=0$, then  $\tilde{C}_4\to 0$.}

\section{Choice of bulk and boundary}

We would like to select as bulk space the four-dimensional space $
(x, x_\pm, z) $ obtained by Son \cite{Son}:\footnote{In the general
setting of  \cite{Son} the space is $(d+3)$-dimensional.} \beq
  ds^2 = - \frac{2 (dx_+)^2}{z^4} +
       \frac{ -2 dx_+ dx_- + (dx)^2 + dz^2 }
            { z^2 }.
  \label{metric}
\eeq

We require that the Schr\"odinger algebra is an isometry  of the
above metric. Here the variable $z$ is the main variable
distinguishing the bulk, namely, the boundary is obtained when
$z=0$. We also need to replace the central element $M$ by the
derivative of the variable $x_-$ which is chosen so that
$\frac{\partial}{\partial x_-}$ continues to be central.   Thus, the
vector-field realization of the Schr\"odinger algebra now becomes:
\bea
  & & H = \frac{\partial}{\partial x_+}, \qquad
      P = \frac{\partial}{\partial x}, \qquad
      M = \frac{\partial}{\partial x_-},
   \nn \\[5pt]
  & & G = x_+ \frac{\partial}{\partial x} + x \frac{\partial}{\partial x_-},
   \label{VecF1} \\[5pt]
  & & D = x \frac{\partial}{\partial x} + z \frac{\partial}{\partial z}
        + 2 x_+ \frac{\partial}{\partial x_+},
    \nn \\[5pt]
  & & K = x_+ \left(
         x \frac{\partial}{\partial x} + z \frac{\partial}{\partial z} + x_+ \frac{\partial}{\partial x_+}
           \right)
        +
        \frac{1}{2} (x^2 + z^2) \frac{\partial}{\partial x_-}. \nn
\eea and it generates an isometry of \eqref{metric}. This
vector-field realization of the Schroedinger algebra acts on the
bulk fields $\phi(x_\pm,x,z) $.

In this realization the Casimir becomes: \bea
\tilde{C}_4 &=& M^2 C_4, \nn \\[5pt]
  C_4 &=&  \hat{Z}^2 - 4 \hat{Z} - 4 z^2  \hat{S} \nn\\[5pt]
  &=& 4 z^2 \del{z}^2 - 8 z \del{z} + 5 - 4 z^2 \hat{S}\ ,
 \label{CasReal2} \\[5pt]
  && \hat{S} \equiv 2 \del{-} \del{+}  - \del{x}^2  \ , \label{schr-op}\\[5pt]
  &&\hat{Z} \equiv 2 z \del{z} - 1.
\eea  Note that \eqref{schr-op} is the pro-Schr\"odinger operator.

Next we consider a realization of the Schr\"odinger algebra on the
boundary. Actually, we use a well known such vector-field
realization \cite{BarRac} in which we only modify the expression for
$M$:
  \bea
  & & H = \frac{\partial}{\partial x_+}, \qquad
      P = \frac{\partial}{\partial x}, \qquad
      M = \frac{\partial}{\partial x_-},
      \nn \\[5pt]
  & & G = x_+ \frac{\partial}{\partial x} + x M ,
   \label{VecF0} \\[5pt]
  & & D = x \frac{\partial}{\partial x} +  \D
        + 2 x_+ \frac{\partial}{\partial x_+},
    \nn \\[5pt]
  & & K = x_+ \left(
         x \frac{\partial}{\partial x} + \D  + x_+ \frac{\partial}{\partial x_+}
           \right)
        +
        \frac{1}{2} x^2 M.  \nn
\eea where   $\D$ is the conformal weight. This vector-field
realization of the Schroedinger algebra acts on the boundary field
$\phi(x_\pm,x) $ with fixed conformal weight $\D$.

In this realization the Casimir becomes:
\bea \tilde{C}^0_4 &=& M^2 C^0_4, \nn \\
  C^0_4 &=&   (2 \Delta - 1) (2 \Delta - 5)
 \label{CasBou} \eea
As expected   $ C^0_4 $ is a constant which has the same value if we
replace $\D$ by $3-\D$: \beq C^0_4 (\D) = C^0_4 (3-\D)
\label{pequi}\eeq  This already means that the two boundary fields
with conformal weights $\D$ and $3-\D$ are related, or in
mathematical language, that the corresponding representations are
(partially) equivalent. This will be very important also below.

\section{Boundary-to-bulk  correspondence}

As we explained in the Introduction we first concentrate on one
aspect of    AdS/CFT \cite{GuKlPo,Witten}, namely, the holography
principle, or boundary-to-bulk correspondence,  which means to have
an operator which maps a boundary field $\varphi$ to a bulk field
$\phi$,    cf.   \cite{Witten}, also
\cite{Dobrev}.\footnote{Mathematically, this means the following. We
treat both the boundary fields and the bulk fields as representation
spaces of the Schr\"odinger algebra. The action of the Schr\"odinger
algebra in the boundary, resp. bulk, representation spaces is given
by formulae \eqref{VecF0}, resp. by formulae  \eqref{VecF1}. The
boundary-to-bulk operator maps the boundary  representation space to
the   bulk representation space.}
%%% NA Aug 25
This will be done within the framework of %%%%
 representation theory without specifying any action.

The fields on the boundary are fixed by the value of the conformal
weight $\D$, correspondingly, as we saw, the Casimir has the
eigenvalue determined by $\D$: \beq C^0_4 \varphi(x_\pm,x) =
\lambda\varphi(x_\pm,x) \ , \qquad \lambda = (2 \Delta - 1) (2
\Delta - 5) \eeq

Thus, the first requirement for the corresponding field on the bulk
$\phi(x_\pm,x,z) $ is to satisfy the same eigenvalue equation,
namely, we require:   \beq C_4 \phi(x_\pm,x,z) =
\lambda\phi(x_\pm,x,z) \ , \qquad \lambda = (2 \Delta - 1) (2 \Delta
- 5) \label{eigen}\eeq where $C_4$ is the differential operator
given in \eqref{CasReal2}. Thus, in the bulk the eigenvalue
condition is a differential equation.

The other condition is the behaviour of the bulk field when we
approach the boundary: \beq
   \phi(x_\pm,x,z) \ \rightarrow \ z^\a
   \varphi(x_\pm,x) \ , \qquad \a = \D,3-\D
   \label{BouBeh}
 \eeq

To find the boundary-to-bulk operator we follow the method of
\cite{DMPPT}, namely, we find the two-point Green function in the
bulk solving the differential equation: \beq
   (C_4 - \lambda)\, G(\chi, z \,;\, \chi', z')
   =
   z{'}^4 \, \delta^3(\chi - \chi') \, \delta(z-z'),
   \label{Gdef}
\eeq where $ \chi = (x_+, x_-, x). $

As in \cite{DMPPT} it is important to use an invariant variable
which in our case is:
\beq
  u = \frac{4 z z'}{ (x-x')^2 - 2(x_+ - x_{+}{'})(x_- - x_{-}{'}) + (z+z')^2
  }.
  \label{udef}
\eeq
 In terms of $u$ the Casimir becomes: \beq
   C_4  = 4 u^2 (1-u) \frac{d^2}{du^2} - 8 u \frac{d}{du} + 5.
   \label{Casimirinu}
\eeq

We can reduce the eigenvalue equation to the equation for the
hypergeometric function by the substitution: $ G(\chi, z \,;\,
\chi', z') = G(u) = u^{\alpha} F(u)$. Then the equation becomes:
\bea
  & & (C_4 - \lambda)\, G(\chi, z \,;\, \chi', z')
  \nn \\
  & & \qquad
  =
  4u^{\alpha+1}
  \left\{
    u(1-u) F'' + 2 (\alpha - 1 - \alpha u) F' +
    \left(
      \frac{ 4 \alpha (\alpha - 3) + 5 - \lambda}{ 4u} - \alpha (\alpha-1)
    \right) F
  \right\} = 0,
  \nn \\
  & & \label{HG1}
\eea where we ignore for the moment the $\delta$-function - it will
be reproduced by the singularity of the solutions at $u=1$. The
parameter $\alpha$ is arbitrary, so we fix it by requiring the
vanishing of the $ u^{-1} $ term, and we recover the two choices:
~$\alpha=\D$, $\alpha=3-\D$. Then we have: \bea
  & & u(1-u) F'' + 2 (\Delta-1 - \Delta u) F'  - \Delta (\Delta - 1) F = 0,
      \quad ( \alpha = \Delta),
     \label{HG2} \\[8pt]
  & & u(1-u) F'' + 2 (2-\Delta - (3-\Delta)u) F' - (\Delta - 2) (\Delta - 3) F = 0,
      \quad ( \alpha = 3-\Delta ).
     \nn \\
  & &   \label{HG3}
\eea  Since the hypergeometric equation has two independent solution,
then it turns out (expectedly) that overall for the function $G(u)$
we also have a single set of two solutions: \bea
  & & G(u) = u^{\Delta} F(\Delta,\Delta-1;2(\Delta-1);u), \label{G1} \\[5pt]
  & & G(u) = u^{3-\Delta} F(3-\Delta,2-\Delta;2(2-\Delta);u). \label{G2}
\eea where $F =~ _2F_1$ is the standard hypergeometric function.

As expected at $u=1$ both solutions are singular:
 by \cite{BaEr}, (\ref{G1}) reads:
  $$
     G(u) = \frac{ u^{\Delta} }{ 1-u } F(\Delta-2,\Delta-1;2(\Delta-1);u),
  $$
 while (\ref{G2})      reads:
  $$
    G(u) = \frac{u^{3-\Delta}}{1-u} F(1-\Delta,2-\Delta;2(2-\Delta);u).
  $$

Following the general method the boundary-to-bulk operator is
obtained from the two-point bulk Green function by bringing one of
the points to the boundary, however, one has to take into account
all info from the field on the boundary. More precisely, in
mathematical terms we express the function in the bulk with boundary
behaviour \eqref{BouBeh} through the function on the boundary by the
formula: \beq
  \phi(\chi,z) = \int d^3\chi' \, S_\a(\chi-\chi',z) \, \varphi(\chi'),
  \label{Bo2Bu}
\eeq where $ d^3 \chi' = dx_+{'} dx_-{'} dx' $ and  $
S_\a(\chi-\chi',z) $ is defined  by \beq
   S_\a(\chi-\chi',z) = \lim_{z' \rightarrow 0} z'{}^{-\a} \, G(u)
   = \left[
        \frac{ 4z }{ (x-x')^2 - 2(x_+-x_+{'}) (x_- - x_-{'}) + z^2 }
     \right]^{\a}.
   \label{Sdef}
\eeq where ~$\a$~ is as in \eqref{BouBeh}.

\section{Intertwining properties}

An important ingredient of our approach is that the bulk-to-boundary
and boundary-to-bulk operators are actually intertwining
operators.\footnote{For the relativistic AdS/CFT case this was done
in \cite{Dobrev}.} To see this we need some more notation.

Let us denote by $L_\a$ the bulk-to-boundary operator : \beq (L_\a \
\phi) (\chi) \doteq \lim_{z \rightarrow 0}z^{-\a}\phi(\chi,z),
\label{Buboug} \eeq where ~$\a = \D,3-\D$ ~consistently with
\eqref{BouBeh}. The intertwining property is: \beq L_\a\circ \hX =
\tX_\a \circ L_\a , \qquad X\in\sch{1}, \label{inta} \eeq where
$\tX_\a$ denotes the action of the generator $X$ on the boundary
\eqref{VecF0} (with $\D$ replaced by $\a$ from \eqref{BouBeh}),
$\hX$ denotes the action of the generator $X$ in the bulk
\eqref{VecF1}. Checking \eqref{inta} is straightforward.

Let us denote by $\tL_\a$ the boundary-to-bulk operator in
\eqref{Bo2Bu}: \beq \phi(\chi,z) = ( \tL_\a \varphi) (\chi,z)
\doteq
 \int d^3\chi' \, S_\a(\chi-\chi',z) \, \varphi(\chi'). \label{bobuop}\eeq
The intertwining property now is: \beq \tL_\a\circ \tX_{3-\a} = \hX
\circ \tL_\a , \qquad X\in\sch{1}. \label{intb} \eeq    The checking
of \eqref{intb} requires some work, but is straightforward.

Next we check consistency of the bulk-to-boundary and
boundary-to-bulk operators, namely, their consecutive application in
both orders should be the identity map: \bea L_{3-\a} \circ \tL_{\a}
&=& {\bf 1}_{\rm boundary}, \label{boubu}\\
\tL_{\a} \circ L_{3-\a} &=& {\bf 1}_{\rm bulk}. \label{bubou}\eea

Checking \eqref{boubu} means: \bea\label{invc} (L_{3-\a} \circ
\tL_{\a}\,\varphi) (\chi) &=& \lim_{z \rightarrow 0} z{}^{\a-3}
\,(\tL_{\a}\,\varphi)\, (\chi,z)  \nn\\ &=&\lim_{z \rightarrow 0}
z{}^{\a-3}
\int d^3\chi' \, S_{\a}(\chi-\chi',z) \, \varphi(\chi') \nn\\
&=& \lim_{z \rightarrow 0} z{}^{\a-3} \int d^3\chi' \, \left(
\frac{4z}{A} \right)^{\alpha}\, \varphi(\chi') \ , \nn\\ && A =
(x-x')^2 -2 (x_+-x'_+)(x_--x'_-) + z^2. \nn \eea

For the above calculation we interchange the limit and the
integration, and use the following formula: \beq\label{delf}
   \lim_{z \rightarrow 0} z^{\a-3} \left( \frac{4z}{A} \right)^{\alpha}
   =
   2^{2\alpha} \pi^{3/2} \, \frac{ \Gamma(\alpha-\frac{3}{2}) }{ \Gamma(\alpha) }
   \, \delta^3(\chi-\chi') \ , \qquad \a - 3/2 \notin  \bbz_-
 \eeq
 The Proof of \eqref{delf} is given in the Appendix.

Using \eqref{delf} we obtain: \beq (L_{3-\a} \circ
\tL_{\a}\,\varphi) (\chi) = 2^{2\alpha} \pi^{3/2} \, \frac{
\Gamma(\alpha-\frac{3}{2}) }{ \Gamma(\alpha) }
   \, \varphi (\chi). \eeq Thus, in order to obtain \eqref{boubu}
 exactly, we have to normalize, e.g., $\tL_{\a}$.

 We note the excluded values ~$\a - 3/2 \notin  \bbz_-$~ for which
 the two intertwining operators are not inverse to each other. This
 means that at least one of the representations is reducible.
This reducibility was established \cite{DoDoMr} for the associated
Verma modules with lowest weight determined by the conformal weight
$\D$.\footnote{For more information on the representation theory and
related hierarchies of invariant differential operators and
equations, cf. \cite{ADDMS}.}

 Checking \eqref{bubou} is now straightforward, but also fails  for
 the excluded values.

Note that checking \eqref{boubu} we used \eqref{Buboug} for $\a \to\
3-\a$, i.e., we used one possible limit of the bulk field
\eqref{Bo2Bu}. But it is important to note that this bulk field has
also the boundary as given in \eqref{Buboug}. Namely, we can
consider the field:
 \beq \varphi_0 (\chi) \doteq (L_\a \ \phi) (\chi) =
\lim_{z \rightarrow 0}z^{-\a}\phi(\chi,z), \label{Bubougg} \eeq where
$\phi(\chi,z)$ is given by \eqref{Bo2Bu}. We obtain immediately:
\beq \varphi_0 (\chi) = \int d^3\chi' \, G_\a(\chi-\chi') \,
\varphi(\chi'),
  \label{Bo2Bu2} \eeq
where \beq G_\a(\chi) = \left[
        \frac {4}{ x^2 - 2x_+ x_- }
     \right]^{\a}.
   \label{Gdef2}\eeq
If we denote by $G_\a$ the operator in \eqref{Bo2Bu2} then we have
the intertwining property: \beq \tX_\a \circ G_\a = G_\a \circ
\tX_{3-\a}\  \ . \label{intc} \eeq Thus, the two boundary fields
corresponding to the two limits of the bulk field are equivalent
(partially equivalent for $\a \in  \bbz + 3/2$). The intertwining
kernel has the properties of the conformal two-point function
\cite{Dobrev}.

Thus, for generic $\D$ the bulk fields obtained for the two values
of $\a$ are not only equivalent - they coincide, since both have the
two fields $\varphi_0$ and $\varphi$ as boundaries.

\medskip

\noindent {\bf Remark:} ~~For the relativistic AdS/CFT
correspondence the above analysis relating the two fields in
\eqref{Bo2Bu2} was given in \cite{Dobrev}. An alternative treatment
relating these two fields via the Legendre transform was given in
\cite{KleWit}.

As in the relativistic case there is a range of dimensions when both
fields $\D,3-\D$ are physical: \beq\label{phys} \D_-^0 \equiv 1/2  <
\D < 5/2 \equiv \D_+^0   \ .\eeq  The above bounds are determined by
the values at which the Casimir eigenvalue $\lambda =
(2\D-1)(2\D-5)$ becomes zero.\footnote{Since the Casimir is fixed up
to additive and multiplicative constants, the latter statement
becomes unambiguous by the requirement that ~$\D_-^0 = 3-\D_+^0\,$.}

%\newpage

\section{Nonrelativistic reduction}

  In order to connect our approach with that of previous works
  \cite{Son,BaMcG,FueMor},
we consider  the action for a scalar field in the background
(\ref{metric}): \beq
  I(\phi) = -\int d^3\chi dz \sqrt{-g} \,
   ( \partial^{\mu} \phi^* \del{\mu} \phi + m_0^2 |\phi|^2).
  \label{action1}
\eeq
By integrating by parts,  and taking into account a non-trivial
contribution from the boundary, one can see that  $ I(\phi) $ has
the following expression:
\beq
  I(\phi) = \int d^3\chi dz \sqrt{-g} \, \phi^* (\square - m_0^2) \phi
  - \lim_{z \rightarrow 0}
    \int d^3\chi \frac{1}{z^3} \phi^* \, z \del{z} \phi.
    \label{action2}
\eeq
The second term is evaluated using (\ref{Bo2Bu}).
For $ z \rightarrow 0, $ one has
\beq
  z \del{z} \phi \ \sim \
  \alpha (4z)^{\alpha} \int d^3\chi'
  \frac{ \varphi(\chi') }{ [(x-x')^2 - 2(x_+-x'_+)(x_- - x'_-)]^{\alpha} }
  + O(z^{\alpha+2}).
  \label{zdphi-bou}
\eeq
It follows that
\bea
 & &
  \lim_{z \rightarrow 0}
    \int d^3\chi \frac{1}{z^3} \phi^* \, z \del{z} \phi
   = \lim_{z \rightarrow 0}  \alpha \int d^3\chi d^3 \chi'
    z^{\a-3} \phi^*(\chi,z) \left( \frac{4}{A} \right)^{\alpha} \varphi(\chi')
 \nn \\[5pt]
 & &
  = 4^{\alpha} \alpha \int d^3\chi d^3 \chi'
    \frac{ \varphi(\chi)^* \varphi(\chi') }{ [(x-x')^2 - 2(x_+ - x'_+)(x_- - x'_-) ]^{\alpha} }.
 \label{action3}
\eea

 The equation of motion being read off from the first term
in (\ref{action2}) can be expressed in terms of the
differential operator (\ref{CasReal2}):
\beq
  (\square  - m_0^2) \, \phi
  = \left(  \frac{C_4 - 5}{4}  + 2 \del{-}^2 - m_0^2  \right) \phi = 0.
  \label{EoM}
\eeq The fields in the bulk (\ref{Bo2Bu}) do not solve the equation
of motion. Now we set an Ansatz for the fields on the boundary: $
\varphi(\chi) = e^{M x_-} \varphi(x_+,x) $ and compactify the $x_-$
coordinate:  $ x_- + a \sim x_-. $ This leads to a separation of
variables for the fields in the bulk in the following way:
\[
  \phi(\chi,z) =
  e^{Mx_-}
  \int dx'_+ dx' \int_0^a d\xi
  \left(
    \frac{4z}{ (x-x')^2 - 2(x_+-x'_+) \xi + z^2 }
  \right)^{\alpha} e^{-M\xi} \varphi(x'_+,x').
\]
Thus we are allowed to make the identification $ \del{-} = M $
both in the bulk and on the boundary \cite{Son,BaMcG}.
We remark that under this identification the operator (\ref{schr-op})
becomes the Schr\"odinger operator.
Integration over $ \xi $ turns out to be incomplete gamma function:
\bea
  & & \phi(\chi,z) = e^{Mx_-} \phi(x_+,x,z),
  \label{nonrel-bulk-fun} \\[7pt]
  & & \phi(x_+,x,z)
      =
      (-2z)^{\alpha} M^{\alpha-1}
      \gamma(1-\alpha, Ma)
  \nn \\[5pt]
  & & \qquad \qquad \quad \times
      \int \frac{ dx'_+ dx' }{ (x_+ - x'_+)^{\alpha} }
      \exp\left( - \frac{ (x-x')^2 + z^2 }{ 2(x_+ - x'_+) }M \right) \,
      \varphi(x'_+,x').
   \label{nonrel-bulk-fun2}
\eea This formula was obtained first in \cite{FueMor}. The equation
of motion (\ref{EoM}) now reads \beq
 \left(
   \frac{\lambda - 5}{4} - m^2
 \right) \phi(x_+,x,z) = 0,
\eeq where  $ m^2 = m_0^2 - 2M^2. $  Requiring $ \phi(x_+,x,z) $ to
be a solution to the equation of motion makes the connection between
the conformal weight and mass: \beq
  \Delta_\pm = \frac{1}{2} ( 3 \pm \sqrt{9 + 4m^2} ).
  \label{Delta2mass}
\eeq This result is identical to the relativistic AdS/CFT
correspondence \cite{GuKlPo,Witten}.  The action (\ref{action2})
evaluated for this classical solutions has the following form
($\a=\D_\pm$): \bea
 & &
  I(\phi) =
    -(-2)^{\alpha} M^{\alpha-1} \a \gamma(1-\alpha,Ma)
 \nn \\[5pt]
 & & \qquad \qquad \times
  \int \frac{dx dx_+ dx' dx'_+}{ (x_+ - x'_+)^{\alpha} }
  \exp\left( - \frac{(x-x')^2}{2(x_+ - x'_+)} M \right) \,
  \varphi(x_+,x)^* \varphi(x'_+,x').
  \label{bou-ansatz}
\eea The two-point function of the operator dual to $ \phi $
computed from (\ref{bou-ansatz})  coincides with the result of
\cite{Son,BaMcG,Henkel,StoHen}.
%%% NA Aug 25 from here
%We remark that our solution
%(\ref{nonrel-bulk-fun}) is more general than the ones in
% \cite{Son,BaMcG}.
%If we use the Ansatz $ \varphi(\chi) = \exp(Mx_- -
%\omega x_+ + ikx) $ for the fields on the boundary, then the
%solutions given by modified Bessel functions \cite{Son,BaMcG} are
%recovered.
We remark that the Ansatz for the boundary fields
$ \varphi(\chi) = \exp(Mx_- -\omega x_+ + ikx) $ used in \cite{Son,BaMcG}
is not necessary to derive (\ref{bou-ansatz}).

  One can also recover the solutions in \cite{Son,BaMcG}
rather simply  in our group theoretical context. We use again the
eigenvalue problem of the differential operator (\ref{CasReal2}):
\beq
   C_4 \, \phi(x_+,x,z) = \lambda \, \phi(x_+,x,z). \label{EVprob}
\eeq
but make separation of variables
\medskip
$ \phi(x_+,x,z) = \psi(x_+,x) f(z). $ Then (\ref{EVprob}) is written as follows:
\[
  \frac{1}{f(z)}
  \left(
     \del{z}^2 - \frac{2}{z} \del{z} + \frac{5-\lambda}{4z^2}
  \right)
  f(z)
  =
  \frac{1}{\psi(x_+,x)} \hat{S} \psi(x_+,x) = p^2 \ \mbox{(const)}
\]
Schr\"odinger part is easily solved:
$ \psi(x_+,x) = \exp(-\omega x_+ + i k x) $ which gives
\beq
  p^2 = -2M \omega + k^2. \label{enegy}
\eeq
The equation for $ f(z) $ now becomes
\beq
  \del{z}^2 f(z) - \frac{2}{z}\, \del{z} f(z) +
  \left(
    2M\omega - k^2 - \frac{m^2}{z^2}
  \right) f(z) = 0.
  \label{eqforz}
\eeq This is the equation given in \cite{Son,BaMcG} for $ d=1$. Thus
solutions to equation (\ref{eqforz}) are given by modified Bessel
functions: $
  f_{\pm}(z) = z^{3/2} K_{\pm \nu}(pz)
$ where $ \nu $ is related to the effective mass $m$
\cite{Son,BaMcG}.  In our group theoretic approach one can see its
relation to the eigenvalue of $C_4: $ $ \nu = \sqrt{\lambda+4}/2. $

  We close this section by giving the expression of
(\ref{bou-ansatz}) for the alternate boundary field
$ \varphi_0. $ To this end, we again use the Ansatz
$ \varphi(\chi) = e^{M x_-} \varphi(x_+,x) $ for (\ref{Bo2Bu2}).
Then performing the integration over $ x'_- $
it is immediate to see that:
\beq
 \varphi_0(x,x_+) \sim
 e^{Mx_-} \int \frac{dx' dx'_+}{ (x_+ - x'_+)^{\alpha} }
  \exp\left( - \frac{(x-x')^2}{2(x_+ - x'_+)} M \right) \,
  \varphi(x'_+,x').
  \label{Bo2Bo-NR}
\eeq
One can invert this relation since
$ G_{3-\alpha} \circ G_{\alpha} = 1_{\rm boundary}. $
Substitution of (\ref{Bo2Bo-NR}) and its inverse to (\ref{bou-ansatz})
gives the following expression:
\beq
  I(\phi) \sim
  \int \frac{dx dx_+ dx' dx'_+}{ (x_+ - x'_+)^{3-\alpha} }
  \exp\left( - \frac{(x-x')^2}{2(x_+ - x'_+)} M \right) \,
  \varphi_0(x_+,x)^* \varphi_0(x'_+,x').
  \label{bou-ansatz2}
\eeq

%\newpage

\section*{Acknowledgements}
We are grateful to M. Asano for valuable discussions. Most of the
work on this paper was done during the visit of V.K.D. as Guest
Professor at Osaka Prefecture University. V.K.D. is supported in
part by  Bulgarian NSF grant {\it DO 02-257}.

 \section*{Appendix}

 We here give a proof of \eqref{delf}.
By the definition of gamma function
 \beq
   \frac{1}{A^{\alpha}} = \frac{1}{2^{\alpha} \, \Gamma(\alpha)}
   \int_0^{\infty} d\xi \, \xi^{\alpha-1} e^{-\xi A/2}.
   \label{gamma}
 \eeq
Consider the Fourier transformation of (\ref{gamma}):
 \beq
   \int \frac{e^{-i p\cdot X} }{ A^{\alpha} } \frac{d^3X}{ (2\pi)^{3/2} }
   =
   \frac{1}{2^{\alpha}\, \Gamma(\alpha)}
   \int \int_0^{\infty} \xi^{\alpha-1} e^{-ip\cdot X - \xi A/2} \,
   \frac{d\xi \, d^3X}{ (2\pi)^{3/2} },
   \label{step1}
\eeq
where
\bea
   & & X = x-x', \quad X_\pm = x_\pm - x'_\pm, \quad p\cdot X = p_x X + p_+ X_+ + p_-
   X_-,
   \nn  \\[5pt]
   & &  d^3 X = dX \, dX_+\, dX_-. \nn
\eea
 Integration over $ X $ is the Guass integral and integration
over $ X_+ $ gives a $ \delta$-function:
 \[
   (\ref{step1}) = \frac{1}{2^{\alpha}\, \Gamma(\alpha)} \int \int_0^{\infty}
   \xi^{\alpha-3/2} \, \delta(-p_+ - i \xi X_-) \, e^{-p_x^2/2\xi - ip_- X_- - \xi z^2/2} \,
   d\xi \, dX_-.
 \]
Applying $ \delta(\lambda x) = |\lambda|^{-1} \delta(x) $ one has:
  \bea
   (\ref{step1}) &=& \frac{1}{2^{\alpha}\, \Gamma(\alpha) } \int_0^{\infty}
   \xi^{\alpha-5/2} \, e^{-(p_x^2-2p_+ p_-)/2\xi  - \xi z^2/2} \,
   d\xi
   \nn \\[5pt]
   &=& \frac{1}{ 2^{\alpha-1} \, \Gamma(\alpha) }
       \left( \frac{\sqrt{\rho}}{z} \right)^{\alpha-3/2}
       K_{\alpha-\frac{3}{2}} (\sqrt{\rho} z),
    \nn
 \eea
 where $ \rho = p_x^2 - 2p_+ p_- $ and integral representation of
 Bessel function was used:
 \beq
   \int_0^{\infty} d\xi \, \xi^{c-1} e^{-(a^2 \xi + b^2/\xi)/2} =
   2 \left( \frac{b}{a} \right)^c K_c(ab).
 \eeq
Now we make the inverse Fourier transformation to (\ref{step1}):
 \beq
   \frac{1}{A^{\alpha}} = \frac{2}{ 2^{\alpha} \Gamma(\alpha)  }
   \int \left( \frac{\sqrt{\rho}}{z} \right)^{\alpha-3/2}
   K_{\alpha-\frac{3}{2}} (\sqrt{\rho}z) \, e^{i p\cdot X} \,
   \frac{ dp_x dp_+ dp_- }{ (2\pi)^{3/2} }.
 \eeq
It follows that
 \bea
   \lim_{z \rightarrow 0} z^{\a-3} \left( \frac{4z}{A} \right)^{\alpha}
   &=&
   \lim_{z \rightarrow 0} \frac{2^{\alpha+1}}{\Gamma(\alpha)} \int
    z^{\a-3/2} (\sqrt{\rho})^{\alpha-3/2} K_{\alpha-\frac{3}{2}} (\sqrt{\rho}z)
    e^{i p\cdot X} \,
   \frac{ dp_x dp_+ dp_- }{ (2\pi)^{3/2} }
   \nn \\[5pt]
   &=& 2^{2\alpha-3/2} \frac{ \Gamma(\alpha-\frac{3}{2}) }{ \Gamma(\alpha) }
   \int   e^{i p\cdot X} \,
   \frac{ dp_x dp_+ dp_- }{ (2\pi)^{3/2} }
   \nn \\[5pt]
   &=& 2^{2\alpha} \pi^{3/2} \, \frac{ \Gamma(\alpha-\frac{3}{2}) }{ \Gamma(\alpha) }
      \, \delta^3(X).
 \eea
 where we used
 \[
   \lim_{z \rightarrow 0} K_{\nu}(z) \sim \frac{2^{\nu}}{ 2 z^{\nu} } \Gamma(\nu).
 \]
Thus \eqref{delf} has been proved.

%\newpage

\end{document}